\documentclass[10pt,a4paper]{article}

\usepackage{a4wide}

\usepackage{ae} 
\usepackage[T1]{fontenc}
\usepackage[ansinew]{inputenc}
\usepackage{amsmath}
\usepackage{amssymb}

\usepackage{amsfonts}
\usepackage{latexsym,amssymb,amsfonts}
\usepackage{amsthm}
\usepackage{epsfig}
\usepackage{url}
\usepackage{graphicx}  
\usepackage{epstopdf}
\usepackage{subfig}

\newtheorem{thm}{\bf Theorem}[section]
\newtheorem{lem}[thm]{\bf Lemma}
\newtheorem{prop}[thm]{\bf Proposition}

\newtheorem{rem}[thm]{{\bf Remark}}
\newtheorem{df}[thm]{\bf Definition}

\newcommand{\R}{\mathbb{R}}

\begin{document}

\title{On SICA models for HIV transmission\thanks{This is a preprint of the following paper: 
C. J. Silva and D. F. M. Torres, On SICA models for HIV transmission, 
published in Mathematical Modelling and Analysis of Infectious Diseases, 
edited by K. Hattaf and H. Dutta, Springer Nature Switzerland AG. 
Submitted 28/Dec/2019; revised 21/Apr/2020; accepted 24/Apr/2020.}}

\author{Cristiana J. Silva\\ 
\texttt{cjoaosilva@ua.pt}	
\and 
Delfim F. M. Torres\\
\texttt{delfim@ua.pt}}

\date{Center for Research \& Development in Mathematics and Applications (CIDMA)\\ 
Department of Mathematics, University of Aveiro, 3810-193 Aveiro, Portugal}

\maketitle

\begin{abstract}
We revisit the SICA (Susceptible--Infectious--Chronic--AIDS)
mathematical model for transmission dynamics of the human 
immunodeficiency virus (HIV) with varying population size 
in a homogeneously mixing population. We consider SICA  models 
given by systems of ordinary differential equations and some 
generalizations given by systems with fractional and stochastic 
differential operators. Local and global stability results 
are proved for deterministic, fractional, and stochastic-type 
SICA models. Two case studies, in Cape Verde and Morocco, 
are investigated. 
\end{abstract}

\noindent {\bf Keywords:} HIV/AIDS, SICA compartmental models, Deterministic models, 
Fractional models, Stochastic models, Stability analysis, 
Lyapunov functions, Cape Verde and Morocco case studies. 


\section{Introduction}

In this work, we make an overview on SICA compartmental models 
for HIV/AIDS transmission dynamics with varying population size 
in a homogeneously mixing population, given by a system of four equations. 
The SICA model divides the total human population into four mutually-exclusive  
compartments: susceptible individuals ($S$); 
HIV-infected individuals with no clinical symptoms of AIDS 
(the virus is living or developing in the individuals 
but without producing symptoms or only mild ones) 
but able to transmit HIV to other individuals ($I$); 
HIV-infected individuals under antiretroviral therapy (ART), 
the so called  chronic stage with a viral load remaining low ($C$); 
and HIV-infected individuals with AIDS clinical symptoms ($A$).
The total population at time $t$, denoted by $N(t)$, is given by
\begin{equation*}
N(t) = S(t) + I(t) + C(t) + A(t).
\end{equation*} 

The deterministic SICA model was firstly proposed as a sub-model 
of a TB--HIV/AIDS co-infection model and published in 2015, 
see \cite{SilvaTorres:TBHIV:2015}. After, it was generalized 
to fractional \cite{SilvaTorres:frac} and stochastic systems 
of differential equations \cite{Jasmina:Silva:Torres}
and calibrated to the HIV/AIDS epidemic situation in Cape Verde 
\cite{SilvaTorres:EcoComplx,SilvaTorres:PrEP} and Morocco 
\cite{Lotfi:HIV:Morocco:2019}. One of the main goals of SICA models 
is to show that a simple mathematical model can help 
to clarify some of the essential relations between 
epidemiological factors and the overall pattern of the AIDS epidemic 
\cite{MayAnderson:Nature:1987,SilvaTorres:EcoComplx}.

The assumptions of the SICA models are now described.
The susceptible population is increased by the recruitment 
of individuals into the population, assumed susceptible, 
at a rate $\Lambda$. All individuals suffer from natural death, 
at a constant rate $\mu$. Susceptible individuals $S$ acquire HIV infection, 
following effective contact with people infected with HIV, at a rate $\lambda$, 
given by
\begin{equation}
\label{eq:lambda}
\lambda(t) = \frac{\beta}{N(t)} \left( I(t) + \eta_C \, C(t)  + \eta_A A(t)  \right),
\end{equation}
where $\beta$ is the effective contact rate for HIV transmission.
The modification parameter $\eta_A \geq 1$ accounts for the relative
infectiousness of individuals with AIDS symptoms, in comparison to those
infected with HIV with no AIDS symptoms. Individuals with AIDS symptoms
are more infectious than HIV-infected individuals (pre-AIDS) because
they have a higher viral load and there is a positive correlation
between viral load and infectiousness \cite{art:viral:load}.
On the other hand, $\eta_C \leq 1$ translates the partial restoration of immune function
of individuals with HIV infection that use correctly ART \cite{AIDS:chronic:Lancet:2013}.

HIV-infected individuals, with and without AIDS symptoms, have access to ART treatment. 
HIV-infected individuals with no AIDS symptoms, $I$, progress to the class 
of individuals with HIV infection under ART treatment $C$ at a rate $\phi$, 
and HIV-infected individuals with AIDS symptoms are treated for HIV at rate $\gamma$.
An HIV-infected individual with AIDS symptoms, $A$, 
that starts treatment, moves to the class of HIV-infected individuals, $I$, 
and will move to the chronic class, $C$, only if the treatment is maintained. 
HIV-infected individuals with no AIDS symptoms, $I$, that do not take 
ART treatment, progress to the AIDS class, $A$, at rate $\rho$. 
Only HIV-infected individuals with AIDS symptoms $A$ suffer 
from an AIDS induced death, at a rate $d$. 
These assumptions are translated into the following mathematical model, 
given by a system of four ordinary differential equations:
\begin{equation}
\label{eq:model:1}
\begin{cases}
\dot{S}(t) = \Lambda - \beta \lambda(t) S(t) - \mu S(t),\\[0.2 cm]
\dot{I}(t) = \beta \lambda(t) S(t) - (\rho + \phi + \mu)I(t)
+ \gamma A(t)  + \omega C(t), \\[0.2 cm]
\dot{C}(t) = \phi I(t) - (\omega + \mu)C(t),\\[0.2 cm]
\dot{A}(t) =  \rho \, I(t) - (\gamma + \mu + d) A(t).
\end{cases}
\end{equation}

The region
\begin{equation}
\label{inv:region:HIV}
\Omega =  \biggl\{ \left( S, I, C, A \right) \in \R_+^{4} \, : \, N \leq \Lambda/\mu \biggr\}
\end{equation}
is positively invariant and attracting \cite{SilvaTorres:TBHIV:2015}. Thus, the dynamics
of the HIV model evolves in $\Omega$.

Model \eqref{eq:model:1} has a disease-free equilibrium, given by
\begin{equation}
\label{eq:dfe}
\Sigma_0 = \left( S^0, I^0, C^0, A^0  \right) 
= \left(\frac{\Lambda}{\mu},0, 0,0  \right).
\end{equation}
Following the approach from \cite{van:den:Driessche:2002}, the basic reproduction 
number $R_{0}$ for model \eqref{eq:model:1}, which represents 
the expected average number of new HIV infections produced 
by a single HIV-infected individual when in contact with 
a completely susceptible population, is given by
\begin{equation}
\label{eq:R0:model:1}
R_0 = \frac{ \beta\, \left(  \xi_2  \left( \xi_1 +\rho\, \eta_A \right) 
+ \eta_C \,\phi \, \xi_1 \right) }{\mu\, \left(  \xi_2  \left( \rho + \xi_1\right) 
+\phi\, \xi_1 +\rho\,d \right) +\rho\,\omega\,d} 
= \frac{\mathcal{N}}{\mathcal{D}},
\end{equation}
where along all manuscript $\xi_1 = \gamma + \mu + d$, 
$\xi_2 = \omega + \mu$, and 
$\xi_3 = \rho + \phi + \mu$, see \cite{SilvaTorres:PrEP}.

To find conditions for the existence of an equilibrium for which HIV 
is endemic in the population (i.e., at least one of $I^*$, $C^*$ or 
$A^*$ is nonzero), denoted by $\Sigma_+ = \left(S^*, I^*, C^*, A^* \right)$, 
the equations in \eqref{eq:model:1} are solved in terms of the force 
of infection at the steady-state $\lambda^*$, given by
\begin{equation}
\label{eq:lambdaH:ast}
\lambda^* = \frac{\beta \left( I^* + \eta_C \, C^* + \eta_A A^*  \right)}{N^*}.
\end{equation}
Setting the right hand side of the equations of the model to zero, 
and noting that $\lambda = \lambda^*$ at equilibrium gives
\begin{equation}
\label{eq:end:equil:model:1}
S^* = \frac{\Lambda}{\lambda^* + \mu}\, , \quad 
I^* =-\frac{\lambda^* \Lambda \xi_1 \xi_2}{D} \, , \quad 
C^* =- \frac{\phi \lambda^* \Lambda \xi_1}{D} \, , \quad 
A^* = -\frac{\rho_1 \lambda^* \Lambda \xi_2}{D}
\end{equation}
with 
$D = -\left(\lambda^* + \mu\right)\left(\mu \left(\xi_2 (\rho + \xi_1) 
+ \xi_1 \phi + \rho d \right) + \rho \omega d\right)$, we use 
\eqref{eq:end:equil:model:1} in the expression for $\lambda^*$ 
in \eqref{eq:lambdaH:ast} to show that the nonzero (endemic) 
equilibrium of the model satisfies
\begin{equation*}
\lambda^* = -\mu (1-R_0).
\end{equation*}
The force of infection at the steady-state $\lambda^*$ is positive only if $R_0 > 1$. 
Thus, the existence and uniqueness of the endemic equilibrium follows. 

\begin{rem}
The explicit expression of the endemic equilibrium 
$\Sigma_+$ of model \eqref{eq:model:1} is given by
\begin{equation}
\label{eq:EE}
\begin{split}
S^* &= \frac{ \Lambda (\rho d \xi_2 
- \mathcal{D})}{\mu (\rho d \xi_2 - \mathcal{N})}\, , \quad \quad 
I^* = \frac{\Lambda \xi_1 \xi_2 (\mathcal{D} 
-\mathcal{N})}{\mathcal{D} (\rho d \xi_2 -\mathcal{N} )} \, , \\
C^* &= \frac{\Lambda \phi \xi_1 (\mathcal{D} -\mathcal{N})}{\mathcal{D} 
(\rho d \xi_2 -\mathcal{N} )} \, , \quad \quad
A^* = \frac{\Lambda \rho \xi_2 (\mathcal{D} 
-\mathcal{N})}{\mathcal{D} ( \rho d \xi_2 -\mathcal{N} )}.
\end{split}
\end{equation}
\end{rem}

Adding the equations of model \eqref{eq:model:1}, with $d=0$, 
gives $\dot{N} = \Lambda - \mu N$, so that $N \to \frac{\Lambda}{\mu}$ 
as $t \to \infty$. Thus, $\frac{\Lambda}{\mu}$ 
is an upper bound of $N(t)$ provided that $N(0) \leq \frac{\Lambda}{\mu}$. 
Further, if $N(0) > \frac{\Lambda}{\mu}$, then $N(t)$ decreases to this level. 
Using $N = \frac{\Lambda}{\mu}$ in the force of infection 
$\lambda = \frac{\beta}{N} \left( I + \eta_C \, C  + \eta_A  A \right)$ 
gives a limiting (mass action) system (see, e.g., \cite{FAgusto:bovineTB:2011}). 
Then, the force of infection becomes
$$
\lambda = \beta_1 \left( I + \eta_C \, C  + \eta_A  A \right) \, , 
\quad \text{where} \quad \beta_1 = \frac{\beta \mu}{\Lambda}.
$$
Therefore, we have the following model:
\begin{equation}
\label{eq:model:2}
\begin{cases}
\dot{S}(t) = \Lambda - \beta_1 \left( I(t) + \eta_C \, C(t)  
+ \eta_A  A(t) \right) S(t) - \mu S(t),\\[0.2 cm]
\dot{I}(t) =  \beta_1 \left( I(t) + \eta_C \, C(t)  
+ \eta_A  A(t) \right) S(t) - \xi_3 I(t) + \gamma A(t) + \omega C(t), \\[0.2 cm]
\dot{C}(t) = \phi I(t) - \xi_2 C(t),\\[0.2 cm]
\dot{A}(t) =  \rho \, I(t) - \xi_1 A(t),
\end{cases}
\end{equation}
where $\xi_1 = \gamma + \mu$. Here, different mathematical models based 
on \eqref{eq:model:1} and \eqref{eq:model:2} are considered. 

The paper is organized as follows. 
In Section~\ref{sec:frac}, a general fractional SICA model is proposed 
and the uniform stability of the equilibrium points is given. In 
Section~\ref{sec:stochas}, a stochastic environmental noise is introduced 
into the SICA model \eqref{eq:model:2}. Existence and uniqueness 
of a positive global solution is proved and conditions  
for the extinction and persistence in mean of the disease are provided. 
In Section~\ref{sec:determ}, the deterministic model is analyzed, 
proving the uniform persistence of the total population and local and global 
stability of the equilibrium points, through Lyapunov's direct method and 
LaSalle's invariance principle. Lyapunov functions are provided. Then, 
with Section~\ref{sec:casestudies}, two case studies are analyzed, 
being shown that models \eqref{eq:model:1} and \eqref{eq:model:2}, 
after an adequate calibration of the parameters, describe well the HIV/AIDS 
situation in Cape Verde and Morocco from 1987 to 2014 and 1986 to 2015, 
respectively. The paper ends with Section~\ref{sec:conc} of conclusion.


\section{Fractional SICA model}
\label{sec:frac}

Fractional differential equations (FDEs), also known in the literature
as extraordinary differential equations, are a generalization of 
differential equations through the application of fractional calculus, 
that is, the branch of mathematical analysis that studies different 
possibilities of defining differentiation operators of noninteger order \cite{Almeida:Book:2015,George:AML:1995,SilvaTorres:frac}.

FDEs are naturally related to systems with memory, which explains 
their usefulness in most biological systems \cite{Owolabi:2017}.
Indeed, FDEs have been considered in many epidemiological models \cite{SilvaTorres:frac}. 

In this section, we analyze the general fractional SICA model 
and its uniform stability, proved in \cite{SilvaTorres:frac}. 
We first recall some important definitions and results that 
are used in the proofs of the uniform asymptotic stability of the equilibrium points. 


\subsection{Preliminaries: fractional calculus and uniform asymptotic stability}

We begin by recalling the definition of Caputo fractional derivative. 

\begin{df}[See \cite{GJI:GJI529}]
Let $a > 0$, $t > a$, and $\alpha, a, t \in\mathbb{R} $. The Caputo fractional 
derivative of order $\alpha$ of a function $f \in C^n$ is given by
\begin{equation*}
_{a}^{C}D_{t}^{\alpha}f(t)= \dfrac{1}{\Gamma(n-\alpha)} 
\int_{a}^{t}\dfrac{f^{(n)}(\xi)}{(t-\xi)^{\alpha+1-n}}d\xi,
\qquad n-1<\alpha<n \in \mathbb{N}.
\end{equation*}
\end{df}

Let us consider the following general fractional differential 
equation involving the Caputo derivative:
\begin{equation}
\label{CaputoGeneral}
_{a}^{C}D_{t}^{\alpha} x(t)= f(t,x(t)), 
\qquad \alpha \in (0,1),
\end{equation}
subject to a given initial condition $x_0=x(t_0)$.

\begin{df}[See, e.g., \cite{LiChen:Automatica:2009}]
\label{def:eq}
The constant $x^*$ is an equilibrium point of the Caputo fractional 
dynamic system \eqref{CaputoGeneral} if, and only if, $f(t, x^*) = 0$. 
\end{df}

We recall an extension of the celebrated Lyapunov direct method
for Caputo type fractional order nonlinear systems \cite{Delavari2012}. 

\begin{thm}[Uniform Asymptotic Stability \cite{Delavari2012}]
\label{uniform_stability}
Let $x^*$ be an equilibrium point for the nonautonomous fractional 
order system \eqref{CaputoGeneral} and $\Omega \subset \mathbb{R}^{n}$ 
be a domain containing $x^*$. Let $L:[0,\infty) \times \Omega 
\rightarrow \mathbb{R}$ be a continuously differentiable function
such that
\begin{equation*}
W_1(x) \leq L(t,x(t)) \leq W_2(x)
\end{equation*}
and
\begin{equation*}
_{a}^{C}D_{t}^{\alpha}  L(t,x(t)) \leq -W_3(x)
\end{equation*}
for all $\alpha\in (0,1)$ and all $x \in \Omega$, 
where $W_1(\cdot)$, $W_2(\cdot)$ and $W_3(\cdot)$ 
are continuous positive definite functions on $\Omega$. 
Then the equilibrium point $x^*$ of system \eqref{CaputoGeneral} 
is uniformly asymptotically stable.
\end{thm} 

A useful lemma is proved in \cite{VargasDeLeon201575}, 
where a Volterra-type Lyapunov function is obtained 
for fractional-order epidemic systems. 

\begin{lem}[See \cite{VargasDeLeon201575}]
\label{lemmaVargas}
Let $x(\cdot)$ be a continuous and differentiable function
with $x(t)\in \mathbb{R_{+}}$. Then, for any time instant 
$t \geq t_0$, one has
\begin{equation*}
_{t_{0}}^{C}D_{t}^{\alpha}\left[ x(t)-x^{*}-x^{*}
\ln\dfrac{x(t)}{x^{*}}\right]
\leq \left( 1-\dfrac{x^{*}}{x(t)}\right) \,
{_{t_{0}}^{C}D}_{t}^{\alpha} x(t), 
\qquad x^{*} \in \mathbb{R}^{+}, 
\qquad \forall\alpha\in (0,1).
\end{equation*}
\end{lem}


\subsection{Fractional SICA model: local and uniform stability analysis}

Let us consider the Caputo fractional order SICA epidemiological model
for HIV/AIDS transmission with constant recruitment rate, 
mass action incidence, and variable population size, 
firstly proposed in \cite{SilvaTorres:frac}:
\begin{equation}
\label{mod:frac}
\begin{cases}
_{t_{0}}^{C}D_{t}^{\alpha}S(t) 
= \Lambda - \beta \left( I(t) + \eta_C \, C(t)  
+ \eta_A  A(t) \right) S(t) - \mu S(t),\\[0.2 cm]
_{t_{0}}^{C}D_{t}^{\alpha}I(t) 
= \beta \left( I(t) + \eta_C \, C(t)  + \eta_A  A(t) \right) S(t) 
- \xi_3 I(t) + \omega C(t) + \gamma A(t), \\[0.2 cm]
_{t_{0}}^{C}D_{t}^{\alpha}C(t) = \phi I(t) - \xi_2 C(t),\\[0.2 cm]
_{t_{0}}^{C}D_{t}^{\alpha}A(t) = \rho \, I(t) - \xi_1 A(t).
\end{cases}
\end{equation}

The local asymptotic stability of the disease free equilibrium 
$\Sigma_0$ \eqref{eq:dfe}, comes straightforward from \cite{Matignon:1996} 
and \cite{Ahmed2007}. Stronger stability results are stated next.  

\begin{thm}[See \cite{SilvaTorres:frac}]
Let $\alpha \in (0, 1)$. The disease free equilibrium $\Sigma_0$ \eqref{eq:dfe}, 
of the fractional order system \eqref{mod:frac}, is uniformly asymptotically stable 
in $\Omega$ \eqref{inv:region:HIV}, whenever \eqref{eq:R0:model:1} satisfies $R_0 < 1$.	
\end{thm}

The uniform asymptotic stability of the disease free equilibrium 
$\Sigma_0$ \eqref{eq:dfe} and endemic equilibrium $\Sigma_+$ \eqref{eq:EE} 
of the fractional order system \eqref{mod:frac} are based on 
Theorem~\ref{uniform_stability} and Lemma~\ref{lemmaVargas}. 
 
\begin{thm}[See \cite{SilvaTorres:frac}]
Let $\alpha \in (0,1)$ and \eqref{eq:R0:model:1} be such that $R_0 > 1$.
Then the unique endemic equilibrium $\Sigma_+$ \eqref{eq:EE} of the 
fractional order system \eqref{mod:frac} is uniformly asymptotically 
stable in the interior of $\Omega$ \eqref{inv:region:HIV}. 
\end{thm}

For the numerical implementation of the fractional derivatives, the implementation of the 
Adams--Bashforth--Moulton scheme is carried out in \cite{SilvaTorres:frac}, which is 
based in the \textsf{Matlab} code \textsf{fde12} by Garrappa \cite{Garrappa}. 
This code implements a predictor-corrector PECE method of Adams--Bashforth--Moulton type, 
as described in \cite{Diethelm:1999}. 


\section{Stochastic SICA model}
\label{sec:stochas}

Here, we begin with the deterministic SICA epidemic model 
for HIV transmission \eqref{eq:model:2}. Then, following
\cite{Grafton:BMB:2005,Gray:SIAM:JAM:2011,Greenhalgh:PhysA:2016,%
Lu:PhysA:2005,Tornatore,ZhaoJiang:ApMathLet:2014},
a stochastic environmental noise is introduced making 
the model biologically more realistic.
Precisely, we consider the model
\begin{equation*}
\begin{cases}
d S (t) = \left[ \Lambda - \beta \left( I(t) + \eta_C \, C(t)
+ \eta_A  A(t) \right) S(t) - \mu S(t) \right] dt,\\[0.2 cm]
d I (t) = \left[ \beta \left( I(t) + \eta_C \, C(t)
+ \eta_A  A(t) \right) S(t) - \xi_3 I(t) + \gamma A(t) + \omega C(t) \right] dt, \\[0.2 cm]
d C (t) = \left[ \phi I(t) - \xi_2 C(t) \right] dt,\\[0.2 cm]
d A (t) = \left[ \rho \, I(t) - \xi_1 A(t) \right] dt. 
\end{cases}
\end{equation*}
Next, the fluctuations in the environment are assumed to manifest themselves 
in the transmission coefficient rate $\beta$, so that $\beta \rightarrow  
\beta + \sigma \dot{B}(t)$, where $B(t)$ is a standard Brownian motion 
with intensity $\sigma^2 > 0$. The stochastic model takes then the following 
form (see \cite{Jasmina:Silva:Torres}):
\begin{equation}
\label{eq:model:stoch}
\begin{cases}
d S (t) = \left[ \Lambda - \beta \left( I(t) + \eta_C \, C(t)
+ \eta_A  A(t) \right) S(t) - \mu S(t) \right] dt\\ 
\qquad\qquad - \sigma \left( I(t) + \eta_C \, C(t)
+ \eta_A  A(t) \right) S(t) dB(t),\\[0.2 cm]
d I (t) = \left[ \beta \left( I(t) + \eta_C \, C(t)
+ \eta_A  A(t) \right) S(t) - \xi_3 I(t) + \gamma A(t) + \omega C(t) \right] dt\\ 
\qquad\qquad + \sigma \left( I(t) + \eta_C \, C(t)  + \eta_A  A(t) \right) S(t) dB(t), \\[0.2 cm]
d C (t) = \left[ \phi I(t) - \xi_2 C(t) \right] dt,\\[0.2 cm]
d A (t) = \left[ \rho \, I(t) - \xi_1 A(t) \right] dt.
\end{cases}
\end{equation}

Let $(\Omega, \mathcal{F}, \{\mathcal{F}\}_{t\geq0}, \mathcal{P})$ 
be a complete probability space with filtration $\{\mathcal{F}\}_{t\geq0}$,
which is right continuous and such that $\mathcal{F}$
contains all $\mathcal{P}$-null sets. The scalar Brownian motion $B(t)$ 
of \eqref{eq:model:stoch} is defined on the given probability space.
Moreover, denote $\R_+^4=\left\{(x_1,x_2,x_3,x_4)|x_i>0, i=\overline{1,4}\right\}$.

The existence and uniqueness of a positive global solution of model 
\eqref{eq:model:stoch} is easily proved using similar arguments 
as the ones used in \cite{Lahrouz:2013}. 

\begin{thm}[See \cite{Jasmina:Silva:Torres}]
\label{existence}
For any $t \geq 0$ and any initial value 
$\left( S(0), I(0), C(0), A(0) \right) \in \R_+^4$, 
there is a unique solution $\left(S(t), I(t), C(t), A(t) \right)$ to 
the SDE \eqref{eq:model:stoch} and the solution remains in $\R_+^4$ 
with probability one. Moreover,
$$
N(t) \to \frac{\Lambda}{\mu} \mbox{ as } t \to \infty,
$$
where $N(t)= S(t)+I(t)+ C(t)+ A(t)$.
\end{thm}

The next theorem provides a condition 
for the extinction of the disease.

\begin{thm}[See \cite{Jasmina:Silva:Torres}]
\label{extinction}
Let $Y(t)=\left(S(t), I(t), C(t), A(t) \right)$ be the solution 
of system \eqref{eq:model:stoch} with positive initial value. 
Assume that $\sigma^2>\frac{\beta}{2 \xi_3}$. Then, 
$$
I(t),C(t),A(t)\rightarrow 0 \text{ a.s. and }
S(t)\to \displaystyle \frac{\Lambda}{\mu} \text{ a.s.},
$$
as $t \to +\infty$.
\end{thm}

Let us now recall the notion of persistence in mean.

\begin{df}
\label{persistence}
System \eqref{eq:model:stoch} is said to be persistent in mean if 
$\displaystyle \lim_{t\rightarrow \infty}\frac{1}{t}\int_0^tI(s)ds>0$ a.s.
\end{df}

In what follows, we use the notation 
$$
[I(t)] := \displaystyle \frac{1}{t}\int_0^tI(s)ds.
$$

\begin{thm}
\label{persistence theor} 
Let
\begin{equation}
\label{eq:K1}
K_1=\frac{\beta}{\mu}\left(\frac{\gamma\rho}{\xi_1}-\xi_3
+\frac{\omega\phi}{\xi_2}\right)+\frac{\mu(\xi_1\xi_2- \mu)}{\Lambda(1 
+ \eta_C + \eta_A )}.
\end{equation}
For any initial value $(S(0),I(0), C(0),A(0))\in \R_+^4$ such that 
$$
S(t)+I(t)+ C(t)+ A(t)=N(t) \to \frac{\Lambda}{\mu} \text{ as } t \to \infty, 
$$
if $K_1 \neq 0$, $\frac{1}{K_1}\left(\frac{\Lambda \beta}{\mu}-\xi_1\xi_2 
-\frac{\sigma^2\Lambda^2}{2\mu^2}\right) > 0$ and $\xi_1,\xi_2>1$, then 
the solution $(S(t),I(t), C(t),A(t))$ satisfies
$$
\liminf_{t\rightarrow \infty} [I(t)]
\geq\frac{1}{K_1}\left(\frac{\Lambda \beta}{\mu}
-\xi_1\xi_2 -\frac{\sigma^2\Lambda^2}{2\mu^2}\right).
$$
\end{thm}

Proofs of Theorems~\ref{extinction} and \ref{persistence theor} follow 
the large number theorem for martingales \cite{Gray:SIAM:JAM:2011} and L'H\^opital's rule,
after applying It\^{o}'s formula in an appropriate way.  
For numerical simulations that illustrate  
Theorems~\ref{extinction} and \ref{persistence theor}, 
see \cite{Jasmina:Silva:Torres}. 


\section{Deterministic SICA model}
\label{sec:determ}

Let us first consider model \eqref{eq:model:1}, that we recall here: 
\begin{equation*}
\begin{cases}
\dot{S}(t) = \Lambda - \beta \lambda(t) \frac{\beta}{N(t)} 
\left( I(t) + \eta_C \, C(t)  + \eta_A A(t)  \right) S(t) - \mu S(t),\\[0.2 cm]
\dot{I}(t) = \beta \lambda(t) \frac{\beta}{N(t)} \left( I(t) + \eta_C \, C(t)  
+ \eta_A A(t)  \right) S(t) - \xi_3 I(t)
+ \gamma A(t)  + \omega C(t), \\[0.2 cm]
\dot{C}(t) = \phi I(t) - \xi_2 C(t),\\[0.2 cm]
\dot{A}(t) =  \rho \, I(t) - \xi_1 A(t).
\end{cases}
\end{equation*}

\begin{thm}[See \cite{SilvaTorres:TBHIV:2015}]
\label{theo:persistence:model:1}
The population $N(t)$ is uniformly persistent, that is,
$$
\liminf_{t \to \infty} N(t) \geq \varepsilon 
$$ 
with $\varepsilon > 0$ not depending on the initial data.
\end{thm}

The local and global stability analysis of the disease free 
equilibrium $\Sigma_0$ given by \eqref{eq:dfe} and endemic 
equilibrium point $\Sigma_+$ given by \eqref{eq:EE} is derived 
in \cite{SilvaTorres:TBHIV:2015}. The local asymptotic stability 
of the  disease-free equilibrium, $\Sigma_0$, follows from Theorem~2 
of \cite{van:den:Driessche:2002}, and holds whenever $R_0 < 1$. 

\begin{lem}[See \cite{SilvaTorres:TBHIV:2015}]
The disease free equilibrium $\Sigma_0$ is locally asymptotically stable
if $R_0 < 1$, and unstable if $R_0 > 1$.
\end{lem}

The global asymptotic stability of the disease free equilibrium 
$\Sigma_0$ is proved in \cite{SilvaTorres:TBHIV:2015}, 
following \cite{CChavez:Feng:Huang:2002}.  

\begin{thm}[See \cite{SilvaTorres:TBHIV:2015}]
The disease free equilibrium $\Sigma_0$ is globally 
asymptotically stable for $R_0 < 1$. 
\end{thm}

For the endemic equilibrium point, existence 
and local asymptotic stability holds whenever $R_0 > 1$.

\begin{lem}[See \cite{SilvaTorres:TBHIV:2015}]
The model \eqref{eq:model:1} has a unique endemic equilibrium whenever $R_0 > 1$.
\end{lem}

The local asymptotic stability of the endemic equilibrium
$\Sigma_+$, can be proved using the center manifold theory \cite{Carr:1981},
as described in \cite[Theorem~4.1]{CChavez_Song_2004}. 

\begin{thm}[See \cite{SilvaTorres:TBHIV:2015}]
The endemic equilibrium $\Sigma_+$ is locally asymptotically 
stable for $R_0 \simeq 1$.
\end{thm}

Now, assume that the AIDS-induced death rate can be neglected, i.e., $d=0$, 
and consider the deterministic SICA model given by \eqref{eq:model:2}.
The model \eqref{eq:model:2} has a unique endemic equilibrium 
given by $\tilde{\Sigma}_+ = \Sigma_+|_{d=0}$, 
whenever $\tilde{R}_0 = R_0|_{d=0} > 1$. Defining 
\begin{equation*}
\Omega_0 = \left\{ \left( S, I, C, A \right) \in \Omega \, : \, I=C=A=0 \right\},
\end{equation*}
and considering the Lyapunov function
\begin{equation}
\label{eq:Lyapunov:function}
V = \left(S - S^* \ln(S)\right) + \left(I - I^* \ln(I)\right) 
+ \frac{\omega}{\xi_2} \left(C - C^* \ln(C)\right) 
+ \frac{\gamma}{\xi_1} \left(A - A^* \ln(A)\right) \, , 
\end{equation}
it follows from LaSalle's invariance principle \cite{LaSalle1976} 
that the endemic equilibrium $\Sigma_+$ is globally asymptotically stable. 

\begin{thm}[See \cite{SilvaTorres:PrEP}]
The endemic equilibrium $\tilde{\Sigma}_+$ of model \eqref{eq:model:2} 
is globally asymptotically stable in $\Omega \backslash \Omega_0$ 
whenever $\tilde{R}_0 > 1$.  
\end{thm}


\subsection{General incidence function for $\eta_C = \eta_A = 0$}
\label{subsec:etazero}

In this subsection, we consider a SICA deterministic model with 
a general incidence function $f$, and assume that $\eta_A = \eta_C = 0$. 
The assumption $\eta_A = \eta_C = 0$ is justified by the following arguments: 
\begin{itemize}
\item[--] although individuals in the chronic stage, 
with a low viral load and under ART treatment, can still transmit 
HIV infection, as ART greatly reduces the risk of transmission 
and individuals that take ART treatment correctly are aware 
of their health status, one can assume that individuals in the class $C$ 
do not have risky behaviours for HIV transmission and do not 
transmit HIV virus, i.e., $\eta_C =0$;  
\item[--] individuals with AIDS clinical symptoms, $A$, are responsible 
and do not have any behaviour that can transmit HIV 
infection or, in other cases, are too sick 
to have a risky behavior, i.e., $\eta_A =0$.
\end{itemize}
Mathematically, the assumption $\eta_A = \eta_C = 0$ translates to the
fact that the incidence function $f$ depends only on $S$ and $I$.
Accordingly, we consider the SICA deterministic model with a general 
incidence function $f$ given by (see \cite{Lotfi:HIV:Morocco:2019}) 
\begin{equation}
\label{1}
\begin{cases}
\dot{S}(t) = \Lambda  - \mu S(t)- f\left(S(t),I(t)\right)I(t),\\[0.2 cm]
\dot{I}(t) = f\left(S(t),I(t)\right)I(t) - \xi_3 I(t)
+ \gamma A(t)  + \omega C(t), \\[0.2 cm]
\dot{C}(t) = \phi I(t) - \xi_2 C(t),\\[0.2 cm]
\dot{A}(t) =  \rho \, I(t) - \xi_1 A(t),
\end{cases}
\end{equation}
with initial conditions
\begin{equation}
\label{IC}
S(0)=S_0\geq0, \
I(0)=I_0\geq0, \
C(0)=C_0\geq0, \
A(0)=A_0\geq 0.
\end{equation}
As in \cite{Hattaf1,HattafC,HattafD2,HattafD1,Lotfi1}, the incidence
function $f(S,I)$ is assumed to be non-negative and continuously differentiable
in the interior of $\R^{2}_{+}$. Moreover, we assume the following
hypotheses (\cite{Lotfi:HIV:Morocco:2019}):
\begin{gather}
\label{H1}\tag{$H_{1}$} f(0,I)=0,
\hspace*{0.2 cm} \text{ for all } I \geq 0,\\
\label{H2}\tag{$H_{2}$} \frac{ \partial f}{\partial S}(S,I)> 0,
\hspace*{0.2 cm} \text{ for all } S>0\ \text{ and } \ I \geq 0,\\
\label{H3}\tag{$H_{3}$} \frac{ \partial f}{\partial I}(S,I)
\leq 0 \hspace*{0.2 cm},\hspace*{0.2 cm} \text{ for all }\ S\geq 0\
\text{ and } \ I \geq 0.
\end{gather}
The reason for adopting hypothesis \eqref{H3} is the fact that susceptible individuals
take measures to reduce contagion if the epidemics breaks out.
This idea has first been explored in \cite{MR0529097}.

\begin{thm}[See \cite{Lotfi:HIV:Morocco:2019}]
All solutions of \eqref{1} starting from non-negative initial conditions
\eqref{IC} exist for all $t>0$ and remain bounded and non-negative.
Moreover,
$$
N(t)\leq N(0)+\dfrac{\Lambda}{\mu}.
$$
\end{thm}

The basic reproduction number $R_0$ is given by
\begin{equation}
\label{R0}
R_0=\dfrac{f(\Lambda/\mu,0)\xi_2\xi_1}{\mathcal{D}}. 
\end{equation}

\begin{thm}[See \cite{Lotfi:HIV:Morocco:2019}]
\begin{itemize}
\item[(i)] If $R_0\leq1$, then system \eqref{1} has a unique
disease-free equilibrium $\Sigma_0$ given by \eqref{eq:dfe}.
\item[(ii)] If $R_0>1$, then the disease-free equilibrium $\Sigma_0$ 
given by \eqref{eq:dfe} is still present and 
system \eqref{1} has a unique endemic equilibrium
of the form $E^*=(S^*,I^*,C^*,A^*)$, with $S^*\in\bigg(0,\dfrac{\Lambda}{\mu}\bigg)$, 
$I^*>0$, $C^*>0$, and $A^*>0$.
\end{itemize}
\end{thm}

\begin{thm}[See \cite{Lotfi:HIV:Morocco:2019}]
\label{Ef}
The disease-free equilibrium $\Sigma_0$ given by \eqref{eq:dfe}
is globally asymptotically stable if
$R_0\leq1$.
\end{thm}

The proof of Theorem~\ref{Ef} comes from LaSalle's invariance principle 
\cite{LaSalle1976}, choosing the Lyapunov functional at $\Sigma_0$ as follows:
\begin{equation*}
V_{1}(S,I,C,A)=S-S_{0}-\int_{S_{0}}^{S}\frac{f(S_{0},0)}{f(X,0)}dX
+I+\dfrac{\omega}{\xi_2}C+\dfrac{\gamma}{\xi_1}A,
\end{equation*}
where $S_0=\dfrac{\Lambda}{\mu}$.

Assume now that function $f$ also satisfies the following condition:
\begin{gather}
\label{H4}\tag{$H_{4}$}
\bigg(1-\dfrac{f(S,I)}{f(S,I^*)}\bigg)\bigg(\dfrac{f(S,I^*)}{f(S,I)}
-\dfrac{I}{I^*}\bigg)\leq0, \ \text{ for all } \ S,I>0.
\end{gather}
  
\begin{thm}[See \cite{Lotfi:HIV:Morocco:2019}]
\begin{itemize}
\item[(i)] If $R_0>1$, then the disease-free equilibrium $E_f$ is unstable.
\item[(ii)] If $R_0>1$ and \eqref{H4} holds, then the endemic equilibrium
$E^*$ is globally asymptotically stable.
\end{itemize}
\end{thm}

For any arbitrary equilibrium $\overline{E}= \left(\overline{S},
\overline{I},\overline{C},\overline{A}\right)$,
the characteristic equation is given by
$$
\begin{array}{ccc}
\left|
\begin{array}{cccc}
-\mu-\dfrac{\partial f}{\partial S}\overline{I}-\lambda
& -\dfrac{\partial f}{\partial I}\overline{I}-f(\overline{S},\overline{I})
& 0 & 0 \\ [8 pt]
\dfrac{\partial f}{\partial S}\overline{I}
& \dfrac{\partial f}{\partial I}\overline{I}
+f(\overline{S},\overline{I})-\xi_3 -\lambda
& \omega & \gamma \\[8 pt]
0 & \phi & -(\xi_2 +\lambda) & 0 \\[8 pt]
0 & \rho & 0 & -(\xi_1 +\lambda)
\end{array}
\right| & = & 0. 
\end{array}
$$
Evaluating the characteristic equation at $\Sigma_0$, we have
$$
\lambda^3+a_1\lambda^2+a_2\lambda+a_3=0,
$$
where
\begin{eqnarray*}
a_1 &=& \xi_1+\xi_2+\xi_3- f(S_0,0),\\
a_2 &=& \xi_1\xi_3+\xi_3\xi_2+\xi_2\xi_1
-(\xi_2+\xi_1)f(S_0,0)-\phi \omega-\rho \gamma,\\
a_3 &=& (1-R_0)\mathcal{D}. 
\end{eqnarray*}
It is clear that $a_3<0$ when $R_0>1$. Then,
the disease-free equilibrium $\Sigma_0$ is unstable.

The global stability of the endemic equilibrium $E^*$ 
comes from LaSalle's invariance principle \cite{LaSalle1976}, 
choosing a Lyapunov functional $V_2$ as follows:
\begin{multline*}
V_{2}(S,I,C,A)
=S-S^{*}-\displaystyle{\int_{S^{*}}^{S}}\frac{f(S^{*},I^{*})}{f(X,I^*)}dX
+I-I^{*}-I^{*}\ln\left(\dfrac{I}{I^*}\right)\\
+\dfrac{\omega}{\xi_2}\left(C-C^{*}-C^{*}\ln\left(\dfrac{C}{C^*}\right)\right)
+\dfrac{\gamma}{\xi_1}\left(A-A^{*}-A^{*}\ln\left(\dfrac{A}{A^*}\right)\right).
\end{multline*}

\begin{rem}[See \cite{Lotfi:HIV:Morocco:2019}]
The incidence function $f$ can take many forms. Table~\ref{Table:1} collects
the most popular of such forms that one can find in the existing literature.
For any form of $f(S,I)$ given in Table~\ref{Table:1},
it is easy to verify that $\dfrac{\partial f(S,I)}{\partial \beta}
=\dfrac{f(S,I)}{\beta}$, which is important for examining
the robustness of model \eqref{1} to $\beta$.
\begin{table}[!ht]
\centering
\caption{Some special incidence functions $f(S,I)$,
where $\alpha_i\geq0$, $i=0,\ldots,3$ (see \cite{Lotfi:HIV:Morocco:2019}).}
\label{Table:1}
{\renewcommand{\arraystretch}{2}	
\begin{tabular}{l c c}
\hline\hline
Incidence functions &   ${f(S,I)}$ &  References \\ \hline\hline
Bilinear  & $\beta S$ & \cite{Bilinear1,Bilinear2,Bilinear} \\[0.15cm] \hline
Saturated & $\dfrac{\beta S}{1+\alpha_1S}$ or $\dfrac{\beta S}{1+\alpha_2I}$
& \cite{Saturated1,ZhaoJiang:ApMathLet:2014} \\[0.15cm] \hline
Beddington--DeAngelis & $\dfrac{\beta S }{1+\alpha_1S+\alpha_2I}$
& \cite{Beddington,Cantrell,DeAngelis}\\[0.15cm] \hline
Crowley--Martin & $\dfrac{\beta S}{1+\alpha_1S+\alpha_2I+\alpha_1 \alpha_2SI}$
& \cite{Crowley,Liu,Zhou}\\[0.15cm] \hline
Specific nonlinear  & $\dfrac{\beta S}{1+\alpha_1S+\alpha_2I+\alpha_3SI}$
& \cite{mahrouf1,Hattaf2,HattafD2,Lotfi,Maziane1}\\[0.15cm] \hline
Hattaf--Yousfi & $\dfrac{\beta S}{\alpha_0+\alpha_1S+\alpha_2I+\alpha_3SI}$
& \cite{Hattaf-Yousfi,mahrouf} \\[0.15cm] \hline\hline
\end{tabular}}
\end{table}
\end{rem}

One way to determine the robustness of model \eqref{1} 
to some specific parameter values,
e.g. $\beta$, consists to examine the sensitivity of the basic
reproduction number $R_0$ with respect to such parameter
by the so called \emph{sensitivity index}.

\begin{df}[See \cite{Chitnis,Rodriques}]
\label{sensi}
The normalized forward sensitivity index of a variable $u$,
that depends differentially on a parameter $p$, is defined as
\begin{equation}
\label{sensidf}
\Upsilon^{u}_p := \dfrac{\partial u}{\partial p}\times\dfrac{p}{u}.
\end{equation}
\end{df}

From \eqref{R0} and Definition~\ref{sensi}, we derive the normalized
forward sensitivity index of $R_0$ with respect to $\beta$, using any
form for the incidence function, as the ones found in Table~\ref{Table:1}, 
and we get the following proposition.

\begin{prop}[See \cite{Lotfi:HIV:Morocco:2019}]
The normalized forward sensitivity index of $R_0$ with
respect to $\beta$ is given by
\begin{equation*}
\Upsilon^{R_0}_\beta =\dfrac{\partial f(S_0,0)}{\partial \beta}
\times\dfrac{\beta}{f(S_0,0)}.
\end{equation*}
\end{prop}

\begin{rem}[See \cite{Lotfi:HIV:Morocco:2019,SilvaTorres:EcoComplx}]
The sensitivity index of $R_0$ \eqref{R0} of the model
with respect to $\phi$, $\rho$, $\gamma$
and $\omega$ are given, respectively, by $\Upsilon^{R_0}_\phi
=-\dfrac{\mu \phi \xi_3}{\mathcal{D}}$, $\Upsilon^{R_0}_\gamma
=\dfrac{\rho \gamma(\mu + d)\xi_2}{\mathcal{D} \xi_3}$,
$\Upsilon^{R_0}_\rho =-\dfrac{\rho(\mu+d)\xi_2}{\mathcal{D}}$
and $\Upsilon^{R_0}_\omega =\dfrac{\mu \omega \phi \xi_3}{\mathcal{D}\xi_2}$. 
\end{rem}

\begin{rem}[See \cite{Lotfi:HIV:Morocco:2019}]
For all incidence functions in Table~\ref{Table:1},
$\beta$ is always the most sensitive parameter and
has a high impact on $R_0$. Indeed, $\Upsilon^{R_0}_\beta$
is independent of any parameter of system \eqref{1}
with $\Upsilon^{R_0}_\beta =+1$.
\end{rem}


\section{Two HIV/AIDS case studies: Cape Verde and Morocco}
\label{sec:casestudies}

Two case studies are now given, showing 
that models \eqref{eq:model:1} and \eqref{eq:model:2} 
describe well the HIV/AIDS situations in Cape Verde and Morocco, 
from 1987 to 2014 and 1986 to 2015, respectively. 


\subsection{Cape Verde (1987--2014)}
\label{subsec:capeverde}

In 2014, 409 new HIV cases were reported in Cape Verde, 
accumulating a total of 4,946 cases. Of this total,
1,766 developed AIDS and 1,066 have died.
The municipality with more cases was Praia, 
followed by Santa Catarina (Santiago island) and S\~{a}o Vicente. 
Cape Verde has developed a Strategic National Plan to fight against AIDS, 
which includes ART treatment, monitoring of patients, 
prevention actions, and HIV testing. From the first diagnosis 
of AIDS in 1986, Cape Verde got significant progress in the fight, 
prevention and treatment of HIV/AIDS 
\cite{report:HIV:AIDS:capevert2015,SilvaTorres:EcoComplx}.

\begin{figure}[!htb]
\centering
\includegraphics[scale=0.6]{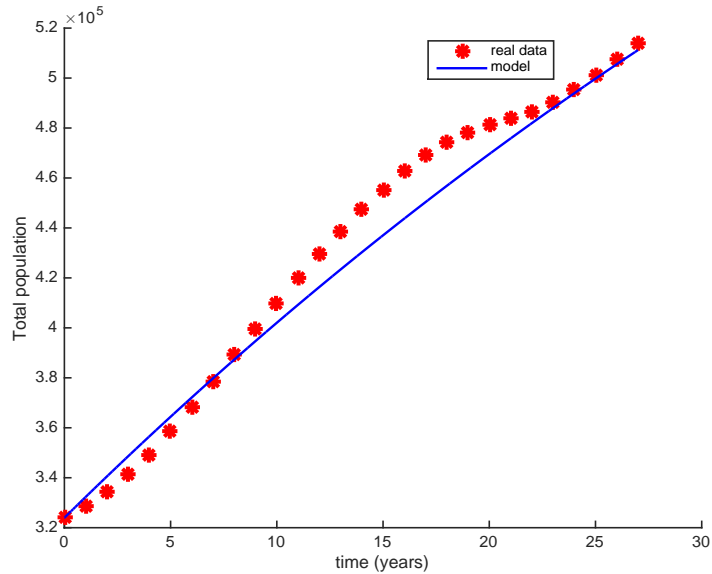}
\caption{Model \eqref{eq:model:1} fitting the total population of Cape Verde between 1987 and 2014 \cite{report:HIV:AIDS:capevert2015,WorldBank:TotalPop:url}.
The $l_2$ norm of the difference between the real total population
of Cape Verde and our prediction gives an error of $1.9\%$ of individuals
per year with respect to the total population of Cape Verde in 2014 (see \cite{SilvaTorres:PrEP}). }
\label{fig:TotalPop}
\end{figure}

\begin{figure}[!htb]
\centering
\subfloat[\footnotesize{$(\beta, \eta_C, \eta_A) = (0.752, 0.015, 1.3)$}]{\label{model:fit:1}
\includegraphics[width=0.48\textwidth]{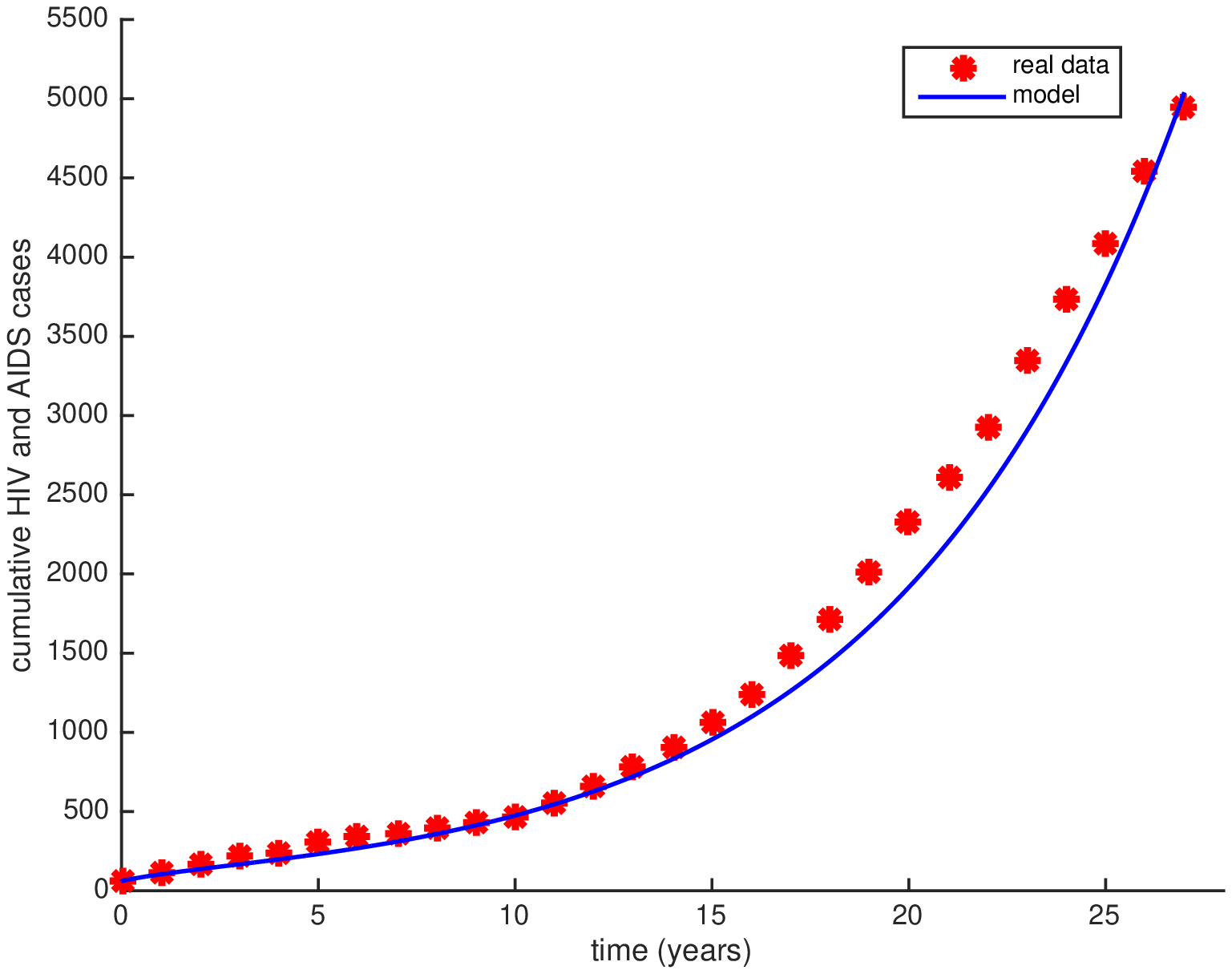}}
\subfloat[\footnotesize{$(\beta, \eta_C, \eta_A) = (0.695, 0.04, 1.35)$}]{\label{model:fit:2}
\includegraphics[width=0.48\textwidth]{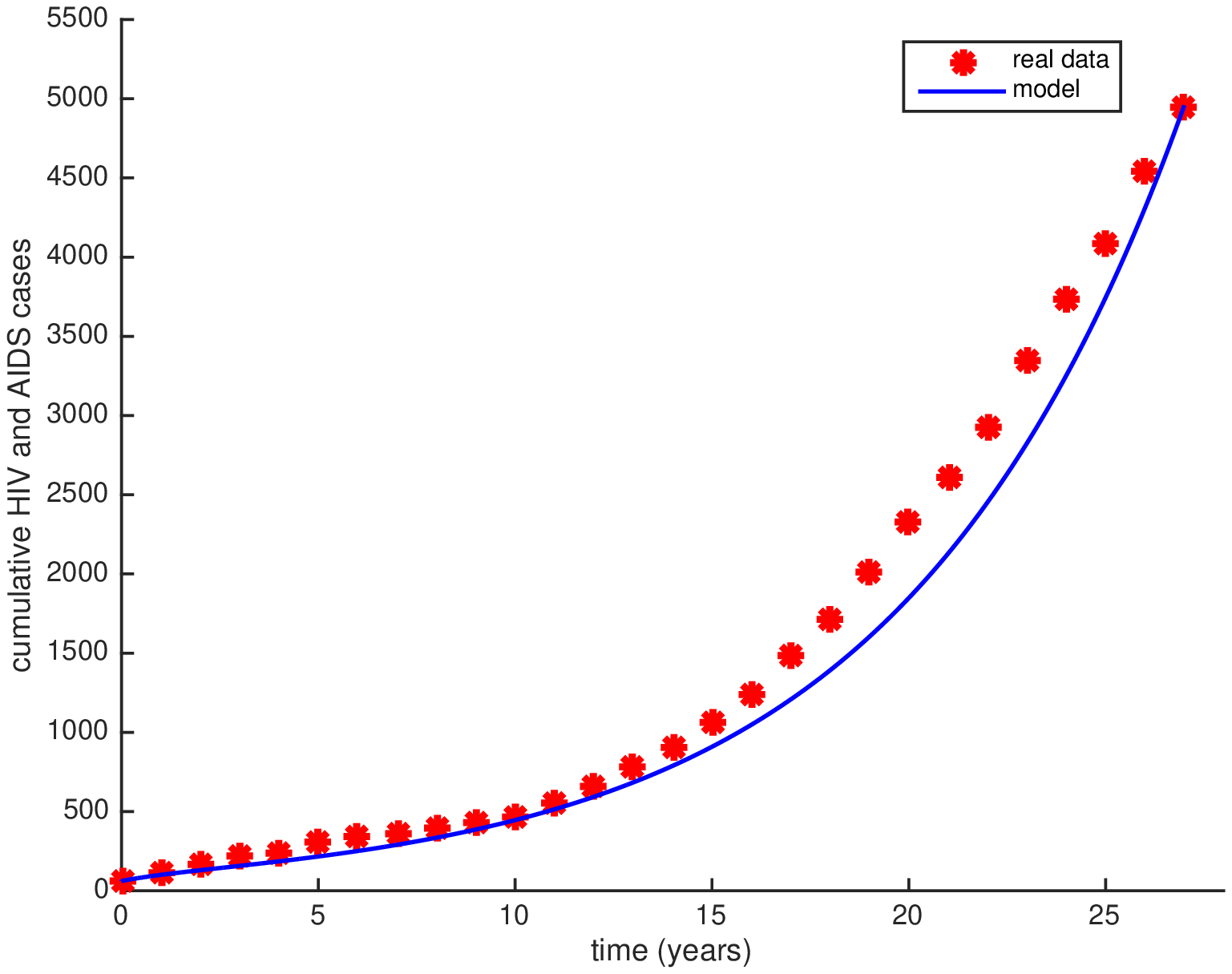}}
\caption{Model \eqref{eq:model:1} fitting the data of cumulative cases of HIV and AIDS infection  
in Cape Verde between 1987 and 2014 \cite{report:HIV:AIDS:capevert2015}. 
The $l_2$ norm of the difference between the real data and the cumulative cases of infection 
by HIV/AIDS given by model \eqref{eq:model:1} gives, in both cases, 
an error of $0.03\%$ of individuals per year with respect 
to the total population of Cape Verde in 2014 (see \cite{SilvaTorres:PrEP}).}
\label{fig:model:fit}
\end{figure}

Model \eqref{eq:model:1} was calibrated to the cumulative cases 
of infection by HIV and AIDS in Cape Verde from 1987 to 2014. 
Following \cite{SilvaTorres:PrEP}, we show that model 
\eqref{eq:model:1} predicts well this reality. 
In Table~\ref{table:realdataCapeVerde}, 
the cumulative cases of infection by HIV and AIDS in Cape Verde 
are depicted for the years 1987--2014 \cite{report:HIV:AIDS:capevert2015}. 
\begin{table}[!htb]
\centering
\caption{Cumulative cases of infection by HIV/AIDS and total population in Cape Verde 
in the period 1987--2014 \cite{report:HIV:AIDS:capevert2015,WorldBank:TotalPop:url}.}
\label{table:realdataCapeVerde}
\begin{tabular}{l l l l l l l l} \hline  \hline
{\small{Year}} & {\small{1987}} & {\small{1988}} & {\small{1989}} & {\small{1990}} 
& {\small{1991}} & {\small{1992}} & {\small{1993}} \\ \hline
{\small{HIV/AIDS}} & {\small{61}} & {\small{107}}  &  {\small{160}} &  {\small{211}} 
&  {\small{244}} & {\small{303}} &  {\small{337}}\\ 
{\small{Population}} & {\small{323972}} & {\small{328861}}  &  {\small{334473}} &  {\small{341256}} 
&  {\small{349326}} & {\small{358473}} &  {\small{368423}}\\ \hline \hline
{\small{Year}} & {\small{1994}} & {\small{1995}} & {\small{1996}} & {\small{1997}} &  {\small{1998}} 
& {\small{1999}} & {\small{2000}} \\ \hline
{\small{HIV/AIDS}} &  {\small{358}} & {\small{395}}  & {\small{432}}		
& {\small{471}} & {\small{560}} & {\small{660}}  &  {\small{779}} \\
{\small{Population}} &  {\small{378763}} & {\small{389156}}  & {\small{399508}}	& {\small{409805}} 
& {\small{419884}}  &  {\small{429576}} &  {\small{438737}}\\ \hline \hline
{\small{Year}} & {\small{2001}} & {\small{2002}} & {\small{2003}} & {\small{2004}} & {\small{2005}}  
& {\small{2006}} & {\small{2007}} \\ \hline
{\small{HIV/AIDS}} &  {\small{913}} & {\small{1064}} &  {\small{1233}} &  {\small{1493}}  &  {\small{1716}}  
& {\small{2015}} & {\small{2334}} \\
{\small{Population}} & {\small{447357}} & {\small{455396}} &  {\small{462675}} &  {\small{468985}} & {\small{474224}}  
& {\small{478265}}	& {\small{481278}} \\ \hline \hline
{\small{Year}} & {\small{2008}} &  {\small{2009}} & {\small{2010}} & {\small{2011}} 
& {\small{2012}} & {\small{2013}} & {\small{2014}}\\ \hline
{\small{HIV/AIDS}} & {\small{2610}} & {\small{2929}} & {\small{3340}}  
&  {\small{3739}} &  {\small{4090}} & {\small{4537}} & {\small{4946}}\\
{\small{Population}} & {\small{483824}}  &  {\small{486673}} &  {\small{490379}} 
&  {\small{495159}} & {\small{500870}} &  {\small{507258}} & {\small{513906}} \\ \hline\hline
\end{tabular}
\end{table}
The values of the initial conditions \eqref{eq:initcond:CV} are based 
on \cite{report:HIV:AIDS:capevert2015,url:worlbank:capevert} (see \cite{SilvaTorres:PrEP}): 
\begin{equation}
\label{eq:initcond:CV}
S_0 = S(0) = 323911 \, , 
\quad I_0 = I(0) = 61\, , 
\quad C_0 = C(0) = 0\, , 
\quad A_0 = A(0) = 0.
\end{equation}
The values of the parameters $\rho = 0.1$ and $\gamma = 0.33$ are taken
from \cite{Sharomi:MathBio:2008} and \cite{Bhunu:BMB:2009:HIV:TB}, respectively.
It is assumed that after one year, the HIV infected individuals $I$ that are under
ART treatment have a low viral load \cite{Perelson} and, therefore,
are transferred to class $C$. In agreement, it is taken $\phi=1$.
It is well known that taking ART therapy is a long-term commitment.
Following \cite{SilvaTorres:EcoComplx}, it is assumed that the default 
treatment rate for $C$ individuals is approximately 11 years ($1/\omega$ years, to be precise).
The AIDS induced death rate is assumed to be $d = 1$ based on \cite{ZwahlenEggerUNAIDS}.
Following the World Bank data \cite{url:worlbank:capevert,WorldBank:TotalPop:url},
the natural death rate is assumed to take the value $\mu = 1/69.54$. 
The recruitment rate $\Lambda = 13045$ was estimated in order to
approximate the values of the total population of Cape Verde
given in Table~\ref{table:realdataCapeVerde}. See 
Figure~\ref{fig:TotalPop}, were it is observable that model 
\eqref{eq:model:1} fits well the total population of Cape Verde.
The AIDS induced death rate is assume to be $d = 1$ based on \cite{ZwahlenEggerUNAIDS}. 
Two cases are considered: $\eta_C = 0.04$, based on a research study known as \emph{HPTN 052}, 
where it is found that the risk of HIV transmission among heterosexual serodiscordant 
couples is 96\% lower when the HIV-positive partner is on treatment \cite{Cohen:NEJM:2011}; 
and $\eta_C = 0.015$, which means that HIV-infected individuals under ART treatment 
have a very low probability of transmitting HIV, based on \cite{DelRomero:2016}. 
For the modification parameter $\eta_A \geq 1$ that accounts for the relative
infectiousness of individuals with AIDS symptoms, in comparison to those
infected with HIV with no AIDS symptoms, we assume $\eta_A = 1.3$ and $\eta_A = 1.35$, 
based in \cite{art:viral:load}. We estimated the value of the HIV transmission rate 
$\beta$ for $(\eta_C, \eta_A) = (0.04, 1.35)$ equal to $0.695$ and for $(\eta_C, \eta_A) = (0.015, 1.3)$ 
equal to $0.752$, and show that model \eqref{eq:model:1} predicts well the reality of Cape Verde 
for these parameter values: see Figure~\ref{fig:model:fit}. 
All the considered parameter values are resumed in Table~\ref{table:parameters:HIV:CV}. 
\begin{table}[!htb]
\centering
\caption{Parameter values of HIV/AIDS models \eqref{eq:model:1} 
and \eqref{1} for Cape Verde \cite{SilvaTorres:PrEP} 
and Morocco~\cite{Lotfi:HIV:Morocco:2019}.}
\label{table:parameters:HIV:CV}
\begin{tabular}{l  p{6.5cm} l l l}
\hline \hline
{\small{Symbol}} &  {\small{Description}} & {\small{Cape Verde}} 
& {\small{Morocco}} & {\small{References}}\\ \hline
{\small{$N(0)$}} & {\small{Initial population}} & {\small{$323 972$}}  & {\small{$23023935$}} & {\small{\cite{url:worlbank:capevert,url:worldbank:morocco}}}\\
{\small{$\Lambda$}} & {\small{Recruitment rate}} & {\small{$13045$}} &  {\small{$2.19\, \mu$}} 
& {\small{\cite{url:worlbank:capevert,url:popdata:morocco} }}\\
{\small{$\mu$}} & {\small{Natural death rate}} & {\small{$1/69.54$}} & $1/74.02$
& {\small{\cite{url:worlbank:capevert,url:popdata:morocco} }}\\
{\small{$\beta$}} & {\small{HIV transmission rate}} & {\small{$0.752$}} & {\small{$0.755$}} & {\small{\cite{SilvaTorres:PrEP,Lotfi:HIV:Morocco:2019}  }}\\
{\small{$\eta_C$}} & {\small{Modification parameter}} & {\small{$0.015$, $0.04$}} & {\small{$0$}}  & {\small{\cite{SilvaTorres:PrEP}}}\\
{\small{$\eta_A$}} & {\small{Modification parameter}} & {\small{$1.3$, $1.35$}} & {\small{$0$}} & {\small{\cite{SilvaTorres:PrEP}}}\\	
{\small{$\phi$}} & {\small{HIV treatment rate for $I$ individuals}} & {\small{$1$}}  
& {\small{$1$}} & {\small{\cite{SilvaTorres:TBHIV:2015,Perelson}}} \\
{\small{$\rho$}} & {\small{Default treatment rate for $I$ individuals}}
& {\small{$0.1 $}} & {\small{$0.1 $}} &{\small{\cite{SilvaTorres:TBHIV:2015,Sharomi:MathBio:2008}}}\\
{\small{$\gamma$}} & {\small{AIDS treatment rate}}
& {\small{$0.33 $}} & {\small{$0.33 $}} &{\small{\cite{SilvaTorres:TBHIV:2015,Bhunu:BMB:2009:HIV:TB}}}\\
{\small{$\omega$}} & {\small{Default treatment rate for $C$ individuals}}
& {\small{$0.09$}} & {\small{$0.09$}} &{\small{\cite{SilvaTorres:TBHIV:2015}}}\\
{\small{$d$}} & {\small{AIDS induced death rate}} & {\small{$1$}}  & {\small{$1$}}
& {\small{\cite{ZwahlenEggerUNAIDS}}}\\
\hline \hline
\end{tabular}
\end{table}

For the triplets $(\beta, \eta_C, \eta_A) = (0.752, 0.015, 1.3)$ 
and $(\beta, \eta_C, \eta_A) = (0.695, 0.04, 1.35)$, and the other parameter 
values from Table~\ref{table:parameters:HIV:CV}, we have that the basic 
reproduction number is given by $R_0 = 4.0983$ and $R_0 = 4.5304$, respectively. 


\subsection{Particular case $\eta_A = \eta_C = 0$, case study in Morocco (1986--2015)}

In \cite{SilvaTorres:EcoComplx} and \cite{Lotfi:HIV:Morocco:2019}, it is assumed that
$\eta_A = \eta_C = 0$, based on the assumptions made in Subsection~\ref{subsec:etazero}. 
Based on these two assumptions, susceptible individuals acquire HIV infection 
by following effective contact with individuals in the class $I$ 
at a rate $\lambda = \beta \frac{I}{N}$.
Taking into account the data from the Ministry of Health 
in Morocco \cite{url:HIVdata:morocco}, in \cite{Lotfi:HIV:Morocco:2019} 
the value of the HIV transmission rate is estimated to be $\beta = 0.755$. 
Moreover, the following initial conditions are considered based on Moroccan data:
\begin{equation}
\label{eq:initcond}
S_0 = (N_0-(2+9))/N_0 ,\quad
\ I_0 = 2/N_0,\quad
\ C_0 = 0,\quad
\ A_0 = 9/N_0,
\end{equation}
with the initial total population $N_0 = 23023935$ \cite{url:worldbank:morocco}.
The values of the parameters $\rho = 0.1$, $\phi=1$, $\omega = 0.09$ and $d = 1$ 
are the same as the ones used in Subsection~\ref{subsec:capeverde}. 
Following \cite{url:popdata:morocco}, the natural death and recruitment rates 
are assumed to take the values $\mu = 1/74.02$ and $\Lambda = 2.19 \mu$. 
All the considered parameter values are summarized 
in Table~\ref{table:parameters:HIV:CV}.

\begin{figure}[!ht]
\advance\leftskip1.6cm
\includegraphics[width=0.7\textwidth]{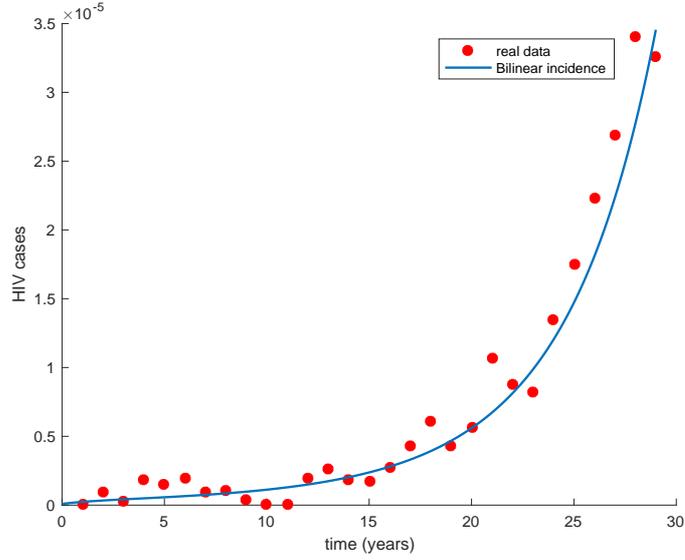}
\caption{Model \eqref{1} fitting the data of HIV cases 
in Morocco between 1986 ($t = 0$) and 2015 ($t = 29$).}
\label{fig:model:fit:Morocco}
\end{figure}

In Figure~\ref{fig:model:fit:Morocco}, we observe that model 
\eqref{1} fits the real data reported in \cite{url:HIVdata:morocco}.
The HIV cases described by model \eqref{1} are given by
$I(t) + C(t) + \mu\left( I(t) + C(t)\right)$
for $t \in [0, 29]$, which corresponds to the interval of time between
the years of 1986 ($t=0$) and 2015 ($t=29$).
	
\begin{figure}[!ht]
\advance\leftskip1.6cm
\includegraphics[width=0.7\textwidth]{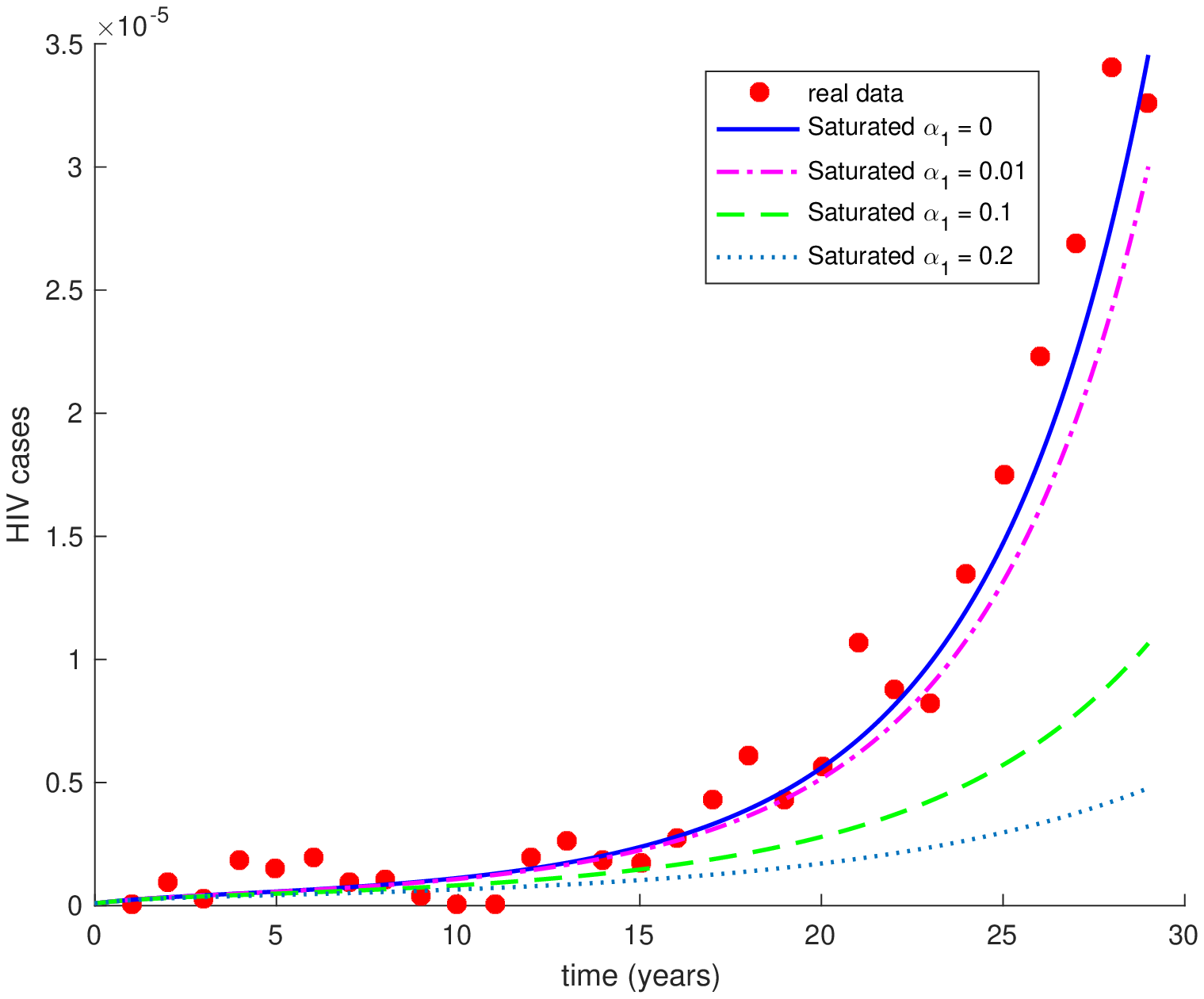}
\caption{Saturated incidence function $\dfrac{\beta S}{1+\alpha_1S}$ 
with $\alpha_1 \in \{0, 0.01, 0.1, 0.2  \}$.
Time $t = 0$ corresponds to the year 1986.}
\label{fig:saturated:alpha1}
\end{figure}

For the saturated incidence function  $\dfrac{\beta S}{1+\alpha_1S}$
\cite{Saturated1}, see Figure~\ref{fig:saturated:alpha1}.
	
\begin{figure}[!ht]
\advance\leftskip1.6cm
\includegraphics[width=0.7\textwidth]{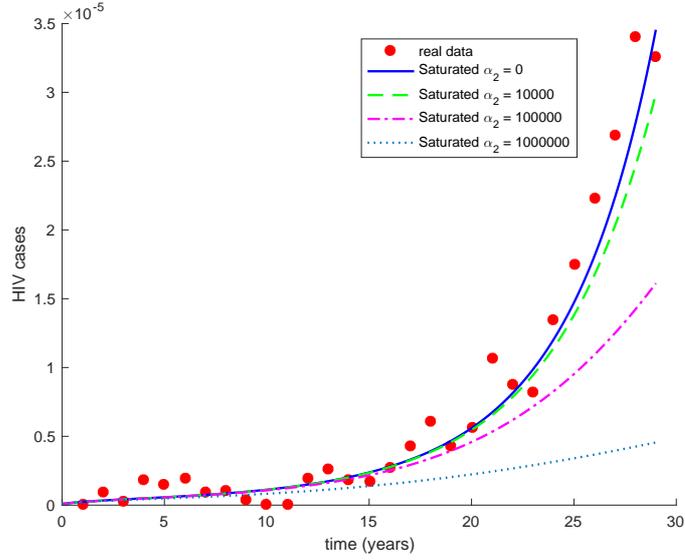}
\caption{Saturated incidence function $\dfrac{\beta S}{1+\alpha_2 I}$ 
with $\alpha_2 \in \{0, 10^4, 10^5, 10^6  \}$.
Time $t = 0$ corresponds to the year 1986.}
\label{fig:saturated:alpha2}
\end{figure}

For the saturated incidence function $\dfrac{\beta S}{1+\alpha_2 I}$
\cite{ZhaoJiang:ApMathLet:2014}, see Figure~\ref{fig:saturated:alpha2}.
	
\begin{figure}[!ht]
\advance\leftskip1.6cm
\includegraphics[width=0.7\textwidth]{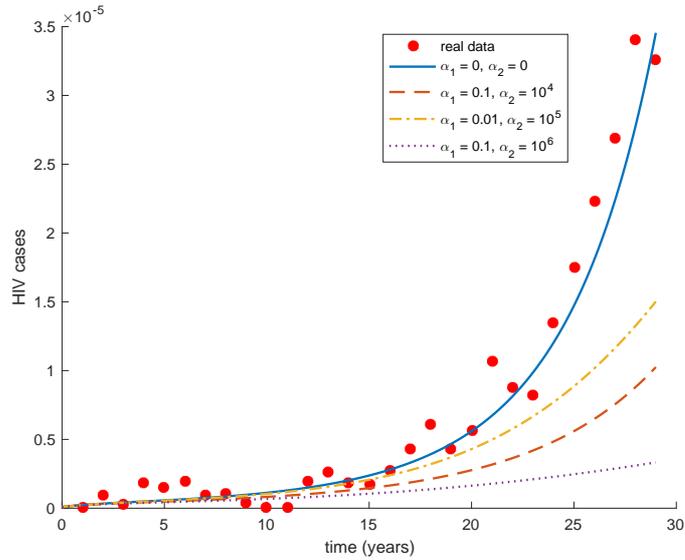}
\caption{Beddington--DeAngelis incidence function 
$\dfrac{\beta S }{1 + \alpha_1 S + \alpha_2 I}$ with 
$\alpha_1 \in \{0, 0.01, 0.1 \}$ and $\alpha_2 \in \{0, 10^4, 10^5, 10^6 \}$.
Time $t = 0$ corresponds to the year 1986, see \cite{Lotfi:HIV:Morocco:2019}.}
\label{fig:Beddington}
\end{figure}

For the Beddington--DeAngelis incidence function
$\dfrac{\beta S }{1 + \alpha_1 S + \alpha_2 I}$
\cite{Beddington,Cantrell,DeAngelis}, see Figure~\ref{fig:Beddington}.
	
\begin{figure}[!ht]
\advance\leftskip1.6cm
\includegraphics[width=0.7\textwidth]{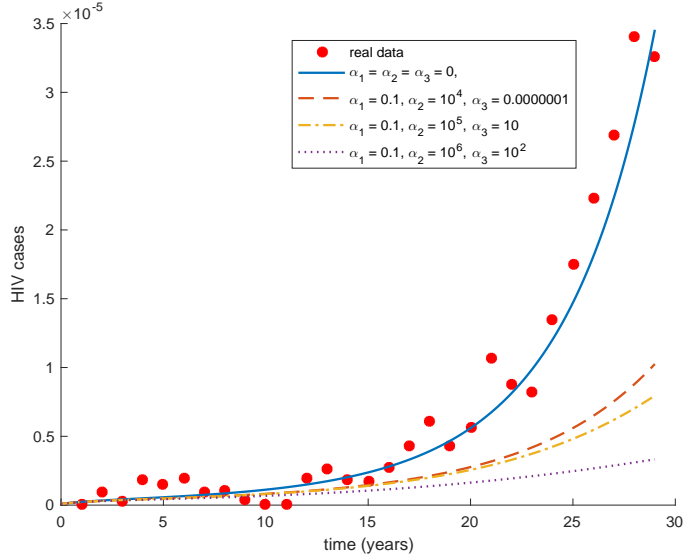}
\caption{Specific non-linear incidence function 
$\dfrac{\beta S}{1+\alpha_1S+\alpha_2I+\alpha_3SI}$ 
with $\alpha_1 \in \{0, 0.1 \}$, $\alpha_2 \in \{0, 10^4, 10^5, 10^6 \}$ 
and $\alpha_3 \in \{0, 10^{-6}, 10, 10^2 \}$. Time $t = 0$ corresponds 
to the year 1986, see \cite{Lotfi:HIV:Morocco:2019}.}
\label{fig:specificNonlinear}
\end{figure}

For the specific non-linear incidence function 
$\dfrac{\beta S}{1+\alpha_1S+\alpha_2I+\alpha_3SI}$
\cite{mahrouf1,Hattaf2,HattafD2,Lotfi,Maziane1}, 
see Figure~\ref{fig:specificNonlinear}.
	
In Tables~\ref{Table:R0} and \ref{Table:R0:alpha1}, the basic reproduction 
number $R_0$ is computed for each incidence function proposed in Table~\ref{Table:1}
with the parameter values from Table~\ref{table:parameters:HIV:CV}, 
see \cite{Lotfi:HIV:Morocco:2019}.
\begin{table}[!ht]
\centering
\caption{Basic reproduction number for some special 
incidence functions, see \cite{Lotfi:HIV:Morocco:2019}.}
\label{Table:R0}
{\renewcommand{\arraystretch}{2}	
\begin{tabular}{l c c}
\hline\hline
Incidence functions &   ${f(S,I)}$ &  $R_0$ \\
\hline\hline
Bilinear  & $\beta S$ & $7.5340$\\[0.15cm]
\hline
Saturated I & $\dfrac{\beta S}{1+\alpha_1S}$
&  $\dfrac{0.0031}{0.0009 \, \alpha_1 + 0.0004}$\\[0.15cm]
\hline
Saturated II &  $\dfrac{\beta S}{1+\alpha_2I}$ &  $7.5340$ \\[0.15cm]
\hline
Beddington--DeAngelis & $\dfrac{\beta S }{1+\alpha_1S+\alpha_2I}$
& $\dfrac{0.0031}{0.0009 \, \alpha_1 + 0.0004}$ \\[0.15cm]
\hline
Crowley--Martin & $\dfrac{\beta S}{1+\alpha_1S+\alpha_2I+\alpha_1 \alpha_2SI}$
& $\dfrac{0.0031}{0.0009 \, \alpha_1 + 0.0004}$\\[0.15cm]
\hline
Specific non-linear  & $\dfrac{\beta S}{1+\alpha_1S+\alpha_2I+\alpha_3SI}$
& $\dfrac{0.0031}{0.0009 \, \alpha_1 + 0.0004}$\\[0.15cm]
\hline
Hattaf--Yousfi & $\dfrac{\beta S}{\alpha_0+\alpha_1S+\alpha_2I+\alpha_3SI}$
& $\dfrac{0.0031}{0.0004 \, \alpha_0 + 0.0009 \, \alpha_1}$ \\[0.15cm]
\hline\hline
\end{tabular}}
\end{table}
\begin{table}[!ht]
\centering
\caption{Basic reproduction number for different values of $\alpha_1$
for the incidence function Saturated~I, Beddington--DeAngelis,
Crowley--Martin and Specific non-linear, see \cite{Lotfi:HIV:Morocco:2019}.}
\label{Table:R0:alpha1}
{\renewcommand{\arraystretch}{2}	
\begin{tabular}{l c c c}
\hline\hline
$\alpha_1$ &   $0.01$ &  $0.1$ & $0.2$ \\
\hline
$R_0$ & $7.3725$ &  $6.1804$ & $5.2392$ \\ [0.15cm]
\hline
\hline
\end{tabular}}
\end{table}
	
In Table~\ref{table:sensitivity:index}, we present the sensitivity index
of parameters $\beta$, $\phi$, $\rho$, $\gamma$ and $\omega$,
computed for the parameter values given in Table~\ref{table:parameters:HIV:CV}.
\begin{table}[!ht]
\centering
\caption{Sensitivity index of $R_0$ for parameter values
given in Table~\ref{table:parameters:HIV:CV} for the bilinear incidence
function $f(S, I) = \beta S$, see \cite{Lotfi:HIV:Morocco:2019}.}
\label{table:sensitivity:index}
{\renewcommand{\arraystretch}{2}
\begin{tabular}{l l }
\hline \hline
{\small{Parameter}} & {\small{Sensitivity index}}\\
\hline
{\small{$\beta$}} & {\small{+1}} \\[-0.2cm]
{\small{$\phi$}} & {\small{$-0.5947$}}  \\[-0.2cm]
{\small{$\rho$}} & {\small{$-0.3437$}}\\[-0.2cm]
{\small{$\gamma$}} &  {\small{$+0.0844$}} \\[-0.2cm]
{\small{$\omega$}} &  {\small{$+0.5170$}} \\
\hline \hline
\end{tabular}}
\end{table}


\section{Conclusion}
\label{sec:conc}

\begin{figure}[!htb]
\centering
\includegraphics[scale=0.6]{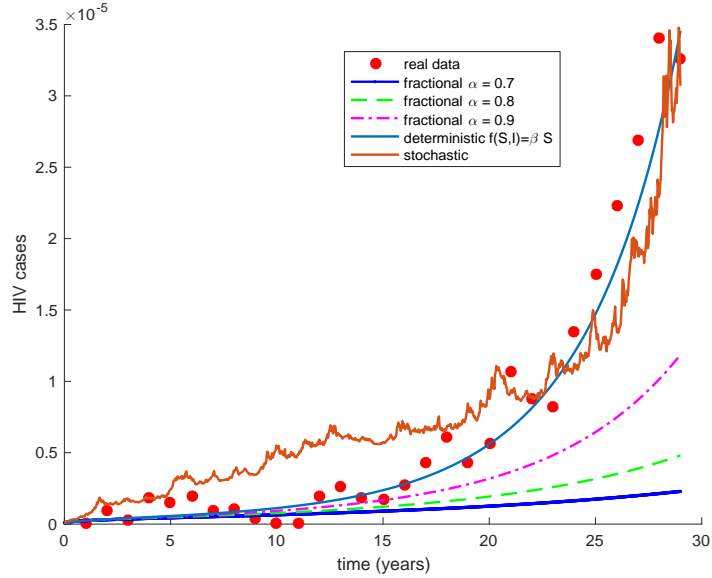}
\caption{Comparison of integer-order, fractional, deterministic 
and stochastic models, \eqref{mod:frac} and \eqref{eq:model:stoch}, 
with $\eta_A = \eta_C = 0$.}
\label{fig:comp}
\end{figure}

We have treated different SICA models
for HIV/AIDS transmission: deterministic and stochastic, 
with integer- and fractional-order derivatives. 
The superiority or better usefulness of one model 
over the others depends always on the concrete situation one is studying.
For example, let us consider the case study of HIV/AIDS infection in Morocco 
from 1986 to 2015, and compare the deterministic model \eqref{mod:frac}, 
both for integer-order ($\alpha = 1$) and fractional-order cases ($\alpha \in (0,1)$), 
with the stochastic model \eqref{eq:model:stoch}. In this case, the results
are shown in Figure~\ref{fig:comp}.
We see that, in this example, the best modeling is obtained using
the deterministic model \eqref{mod:frac} with integer-order derivatives 
($\alpha = 1$). 

All our simulations have been done using the numerical 
computing environment \textsf{Matlab}, release R2016a,
on an Apple MacBook Air Core i5 1.3 GHz with 4Gb RAM. 
The solutions of the models were found in ``real time''.
For example, the total computing time needed to generate and plot 
the 5 solutions in Figure~\ref{fig:comp}, obtained by solving deterministic, 
stochastic, integer-order and fractional models, was as small as 3.03 seconds. 


\section*{Acknowledgements}

The authors were supported by CIDMA through 
the Portuguese Foundation for Science and Technology 
(FCT), within project UIDB/04106/2020.
Silva was also supported by national funds (OE), through FCT, I.P., in the scope 
of the framework contract foreseen in numbers 4, 5 and 6 of art.~23, 
of the Decree-Law 57/2016, of August 29, changed by Law 57/2017, of July 19.
They are sincerely grateful to two anonymous reviewers for
their useful comments.



\end{document}